\documentclass[aps,prstab,groupedaddress,showpacs,showkeys,twocolumn]{revtex4-1}

\usepackage{graphicx} 
\usepackage{url} 
\usepackage{physics} 
\usepackage{amsmath}
\usepackage{color}

\begin{document}

\title{Optimize Nonlinear Beam Dynamical System with Square Matrix Method}

\author{Yongjun Li} \thanks{yli@bnl.gov}

\author{Li Hua Yu}

\author{Lingyun Yang} \thanks{Currently at Renaissance Technologies LLC.}

\affiliation{Brookhaven National Laboratory, Upton, New York 11973}

\begin{abstract}

Nonlinear dynamics has an important role when designing modern synchrotron
lattices. In this letter, we introduce a new method of using a square
matrix to analyze periodic nonlinear dynamical systems
~\cite{Yu:2016,Yu:2010}. Applying the method to the National Synchrotron
Light Source II storage ring lattice has helped to mitigate the chaotic
motion within its dynamic aperture. For a given dynamical system, the
vector space of a square matrix can be separated into different low
dimension invariant subspaces according to their eigenvalues. When Jordan
decomposition is applied to one of the eigenspaces, it yields a set of
accurate action-angle variables. The distortion of the new action-angle
variables provides a measure of the nonlinearity. Our studies show that
the common convention of confining the tune-shift with amplitude to avoid
the crossing of resonance lines may not be absolutely necessary. We
demonstrate that the third order resonance can be almost perfectly
compensated with this technique. The method itself is general, and could
be applied to other nonlinear systems.
\end{abstract}

\maketitle

\section{introduction}
Long-term nonlinear behavior of charged particles in synchrotron plays a
vital role in beam dynamics. To understand the impact nonlinear behavior
has, one can analyze particle motion under many iterations of the
one-turn-map. The reliable numerical approach is using appropriate local
symplectic integration
methods~\cite{Yoshida:1990,Borland:2000,Schmidt:2005}. For the analysis of
the dynamics, however, one can use a more compact representation of the
one-turn-map to extract relevant information. There are many approaches
one can take, such as canonical perturbation theory, Lie operators, power
series, and normal
form\cite{Lichtenberg:1983,Ruth:1986,Guignard:1978,Schoch:1958,
  Dragt:1988,Berz:1989,Chao:2002,Bazzani:1994,Forest:1989,
  Michelotti:1995,Forest:1998,Cai:2011,Wan:2001,Danilov:2010,
  Autin:1985,Hagel:1987}, etc. Here, we would like to study this problem
from a somewhat different perspective (i.e., using linear algebra
techniques.)  The detailed theory on the method has been explained in
ref.~\cite{Yu:2016,Yu:2010}. We will summarize this method in
Section~\ref{theory}, and then describe its applications in
Section~\ref{application}.

\section{theory}\label{theory}

For a given periodic system, such as a particle moving in a synchrotron,
its status can be represented by the complex normalized variable
~\cite{Courant:1997,Dragt:1988,Chao:2002,Lee:1999}
$z=\bar{x}-i\bar{p}=\sqrt{2J}e^{i\psi}$ and its conjugate
$z^{\ast}=\bar{x}+i\bar{p}=\sqrt{2J}e^{-i\psi}$. We use these to form a
truncated vector $\mathbf{Z} =
(1,z,z^{\ast},z^2,zz^{\ast},\cdots,z^{\ast{}n})^T$, where $(J,\psi)$ are
linear action-angle variables, $^T$ is the vector transpose, and $n$ is
the truncated order. The one-turn-map from an initial status
$\mathbf{Z}_0$ to its final status $\mathbf{Z}_1$ is represented by a
square matrix $\mathbf{M}$:
\begin{equation}\label{one_turn_map1}
\mathbf{Z}_1=\mathbf{M}\mathbf{Z}_0.
\end{equation}
The matrix $\mathbf{M}$ is upper-triangular, and has the form 
\begin{equation}\label{matrix}
\mathbf{M} = \left(
\begin{array}{ccccc}
 1 & 0  & \cdots & 0 \\
 0 & \mathbf{M}_{11} & \cdots & \mathbf{M}_{1n} \\
 \vdots & 0 & \ddots  & \vdots \\
 0 & 0  & \cdots & \mathbf{M}_{nn}\\
\end{array}
\right).
\end{equation}
Here different submatrices $\mathbf{M}_{ij}$ have different
dimensions. All diagonal blocks $\mathbf{M}_{ii}$'s are square diagonal
submatrices.

Since the matrix is upper-triangular, its eigenvalues are given by its
diagonal elements in the form of $e^{im\mu}$, where $m$ is an integer, and
$\mu$ is the linear tune. We can separate the full space spanned by the
matrix columns into different invariant subspaces according to the
eigenvalues. We found that the simplest invariant subspace $e^{i\mu}$
already provides a wealth of information about the dynamical system and
the high dimension matrix is reduced to a much lower dimension. For
example, for a $7^{th}$ order 4D phase space system, its original
dimension is $330\times330$. After Jordan decomposition, a set of 4
left-eigenvectors $\mathbf{u}_{k=0,\cdots,3}$ span the invariant subspace
$e^{i\mu}$. A matrix $\mathbf{U}$ consists of these 4 row vectors
satisfies the left-eigenvector equation
\begin{equation}\label{Jordan}
\mathbf{U}\mathbf{M} = e^{i\mu\mathbf{I}+\mathbf{\tau}}\mathbf{U} =
\mathbf{N}\mathbf{U}
\end{equation}
where the $4\times4$ matrix $\mathbf{N}$ is the Jordan block with the
eigenvalue $e^{i\mu}$, corresponding to the $e^{i\mu}$ invariant subspace
inside the space of vector $\mathbf{Z}$. $\mathbf{I}$ is the identity
matrix in this space, while $\mathbf{\tau}$ is a superdiagonal matrix:
\begin{equation}\label{tau}
\mathbf{\tau}=\left(
\begin{array}{ccccc}
0 & 1 &        &   \\
  & 0 & \ddots &   \\
  &   & \ddots & 1 \\
  &   &        & 0 \\
\end{array}
\right).
\end{equation}

The mapping from $\mathbf{Z}_0$ to $\mathbf{Z}_1$ generated by the
one-turn-map $\mathbf{M}$, when projected into this subspace, can be
re-written as
\begin{equation}\label{one_turn_map2}
\mathbf{W}_1\equiv\mathbf{UZ}_1=\mathbf{UMZ}_0=e^{i\mu\mathbf{I}+\mathbf{\tau}}
\mathbf{UZ}_0\equiv{}e^{i\mu{}\mathbf{I}+\mathbf{\tau}}\mathbf{W}_0.
\end{equation}
$\mathbf{W}_0$ can be written as a one-column vector
\begin{equation}\label{invariant}
\mathbf{W}_0^T=(w_0,w_1,w_2,\dots,w_{m-1}),
\end{equation}
where $m$ is the dimension of the invariant subspace. KAM theory states
that the invariant tori are stable under small
perturbation~\cite{Lichtenberg:1983, Bazzani:1994,Poschel:2001}. For
sufficiently small amplitude of oscillation in $\mathbf{Z}$, the invariant
tori are deformed and survive. So the system has a nearly stable frequency
and when the amplitude is small, the fluctuation of the frequency is also
small. Thus for a specific initial condition described by $\mathbf{Z}_0$,
the rotation in the eigenspace should be represented by a phase factor
$e^{i(\mu+\phi)}$ as
\begin{equation}\label{one_turn_map3}
\mathbf{W}_1=e^{i\mu\mathbf{I}+\mathbf{\tau}}\mathbf{W}_0 \cong
e^{i(\mu+\phi)} \mathbf{W}_0.
\end{equation}
 where $\phi$ depends on the initial condition.

$\mathbf{\tau}$ in Eq.~\eqref{tau} has no proper eigenvector, but only
generalized eigenvectors. However, as we increase the order of the Taylor
expansion, the dimension of the eigenspace increases and approaches
infinity, and the eigenvector of $\mathbf{\tau}$ is defined as a coherent
state \cite{Glauber:1963,Sudarshan:1963}:
\begin{equation}\label{coherent}
\mathbf{\tau}\mathbf{W}_0\cong{}i\phi\mathbf{W}_0.
\end{equation}
 The polynomials in
Eq.~\eqref{invariant} are $w_0=u_0\mathbf{Z}_0,
w_1=u_1\mathbf{Z}_0,w_2=u_2\mathbf{Z}_0,\cdots$. Then Eq.~\eqref{coherent}
reads as
\begin{equation}\label{chain}
  \mathbf{\tau}\left(
  \begin{array}{c}
  w_0\\w_1\\ \vdots\\ w_{m-1} \\
   \end{array}\right)
 =
  \left(
\begin{array}{c}
  w_1\\w_2\\ \vdots \\ 0 \\
   \end{array}\right) \cong
\left(\begin{array}{c}
i\phi w_0\\i\phi w_1\\ \vdots \\ i\phi w_{m-1} \\
\end{array}\right).
\end{equation}
When the invariant tori survive and there is a stable frequency, we see
that Eq.~\eqref{chain} requires
\begin{equation}\label{phase}
i\phi=\frac{w_1}{w_0}\cong\frac{w_2}{w_1}\dots\cong\frac{w_{m-1}}{w_{m-2}}.
\end{equation}
Therefore only those vectors $\mathbf{W}_0$ which satisfy
Eq.~\eqref{phase} with $\phi$ as a real number represent a motion with a
stable frequency given by a phase advance $\mu+\phi$ every turn. From
$w_0=u_0\mathbf{Z}_0,\cdots$, we can see that $\phi$ is determined by the
initial value $\mathbf{Z}_0$. $\mu$ represents the zero amplitude tune
while $\phi$ is the amplitude dependent tune-shift. Thus we get a set of
new action-angle variables $(r_j,\theta_j)$
\begin{equation}\label{new_action_angle}
w_{j}=|w_j|e^{i \theta_j}=r_j e^{i \theta_j}, j=0,1,\cdots. 
\end{equation}
Even though all $(r_j,\theta_j)$'s behave like action-angle variables,
they have different power orders of monomials of $z,z^\ast$ , and hence
represent approximation of the action-angle variable to different
precisions. For example, in the case of a $7^{th}$ order square matrix,
$w_0$ has terms of powers from $1^{st}$ to $7^{th}$ order, $w_1$ has terms
of powers from $3^{rd}$ to $7^{th}$ order while $w_3$ has only a small
$7^{th}$ order term $z(zz^\ast)^3$. $w_3$ provides little information
about the rotation in the phase space while $w_0$ has detailed
information. In this paper, we only focus on $w_0$.

A stable motion means the invariant tori can survive with multiple
turns. Applying Eq.~\eqref{one_turn_map3} $n$ times, we obtain
\begin{equation}\label{n_turn_map}
\mathbf{W}_n= e^{in\mu\mathbf{I} +n{\mathbf{\tau}}} \mathbf{W}_0 =
e^{in\mu } e^{n{\tau}} \mathbf{W}_0.
\end{equation}
After a derivation based on Eq.~\eqref{phase} and ~\eqref{n_turn_map}, we
recognize that a stable motion requires (see Eq.(1.19) of ~\cite{Yu:2016})
\begin{equation}\label{stability}
\Im(\phi) \equiv \Im(-\frac{iw_1}{w_0}) \approx 0;
\Delta\equiv\frac{w_2}{w_0}-(\frac{w_1}{w_0})^2 \approx 0.
\end{equation}
We refer to Eq.~\eqref{stability} as the ``coherence conditions'' of a
stable motion. $w_0,\phi$, and $\Delta$ are all functions of the initial
value of $z,z^\ast$. For a given initial value of $|w_0|$, the distortion
of the real part of $\phi$ from a constant is the tune fluctuation, while
the imaginary part of $\phi$ gives “amplitude fluctuation”, i.e., the
variation of $r_0=|w_0|$ after many turns. The non-zero $\Delta$ indicates
a deviation from a coherent state, and it seems to be related to the
Liapunov exponents~\cite{Lichtenberg:1983} and the region of stable
motion.

\section{application}\label{application}
\begin{figure}[!ht]
\centering
\includegraphics[width=1.\columnwidth]{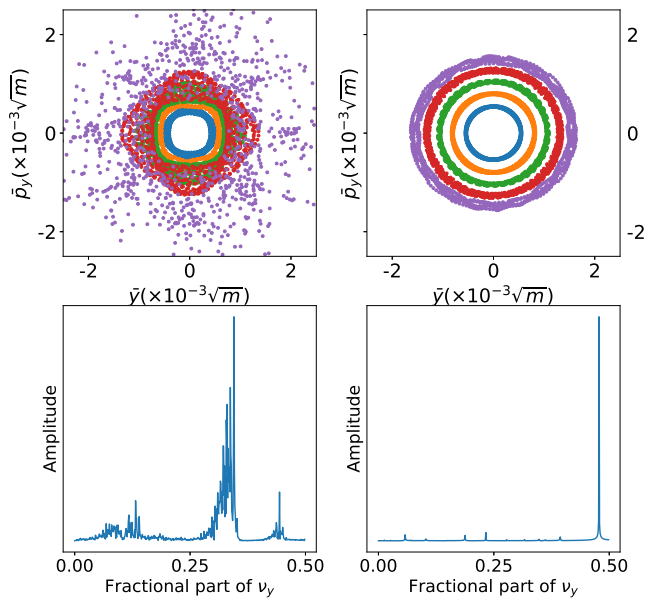}
\caption{\label{phase_y} Comparison of simulated trajectories (top) and
  spectral analysis (bottom) of $\bar{y}-\bar{p}_y$ motion for Solution A
  (left) and Solution B (right). In both plots, 5 pairs of initial
  conditions with the $x$ amplitude gradually increases from 10 to 20$mm$,
  and $y$ increases proportional to $x$ from 1 to 3$mm$. The spectral
  analysis for an initial condition $x=20mm$ and $y=3mm$ also indicates
  that Solution A's motion (bottom, left) is much more chaotic than B
  (bottom, right). The occupied area of Solution A becomes much larger for
  long-term tracking ($>15,000$ turns), but Solution B remains almost the
  same, which indicates the square matrix method is superior in optimizing
  the long term stability.}
\end{figure}

\begin{figure}[!ht]
\includegraphics[width=1\columnwidth]{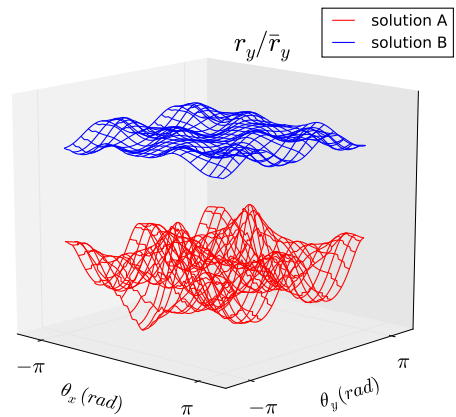}
\caption{\label{flat_plane} Comparison of the distortion of $r_y$ mapped
  from the same constant $J_{x,y}$ planes for both solutions. Solution B
  plane is deliberately shifted up for a clear view.}
\end{figure}

In this Section, we give an example of applying the square matrix method
to optimize a storage ring's dynamic aperture. Consider one particle with
initial linear actions $J_{x,y}$. It is launched for multi-turns
tracking. The linear actions are no longer constants when nonlinearity
dominates over linear dynamics. There is a distortion from flat planes in
the Poincar\'e section. We characterize this distortion by $\Delta J/
J=(J_{max}-J_{min})/J_{mean}$. When the distortion is large, particles
receive large nonlinear kicks and the motion becomes chaotic or even
unstable. The stable region in phase space is defined as dynamic
aperture. The goal of nonlinear optimization is to increase the dynamic
aperture. In the 1D case, this is equivalent to optimizing the
trajectories in the normalized phase space $\bar{x}-\bar{p}_x$ so that
they are as close as possible to circles (see FIG.~\ref{phase_y}, top
right plot). In order to minimize $\Delta J/J$, we need to calculate
$J_{x,y}$ from constant $|w_{x,y}|$, in which an inverse function
calculation is required. There is a way to avoid the inverse function
calculation. Minimizing $\Delta J/J$ is equivalent to optimizing the
system so that constant planes in the Poincar\'e sections in $J_{x,y}$
space are mapped to approximate flat planes in the Poincar\'e sections in
the $|w_{x,y}|$ space (see FIG.~\ref{flat_plane}), and vice
versa. Therefore we map a pair of constant $J_{x,y}$ planes into a pair of
surfaces of $r_{x,y}=|w_{x,y}|$. Then we characterize the nonlinear
distortion by the deviation of surfaces of $r_{x,y}$ from flat planes,
given by
\begin{equation}\label{variation}
\frac{\Delta r}{\bar{r}}=\frac{\Delta
  |w|}{|\bar{w}|}=\frac{r_{max}-r_{min}}{\bar{r}},
\end{equation}
as a measure of nonlinearity. Here $\bar{r}$ is the mean value of $r$. The
system can be optimized by making the surfaces $r_{x,y}$ as close as
possible to constants for various amplitudes.

An application of this method was applied to the National Synchrotron
Light Source-II (NSLS-II), when the lattice had a linear chromaticity of
+7 in both planes. The lattice layout is described in
ref.~\cite{NSLS-II:2013}. After tuning the chromaticity to $+7$ with 3
families of chromatic sextupoles, the optimization knobs were those 6
families of non-chromatic sextupoles. In this case we selected 3 sets of
constant $J_{x,y}$ in the $J_{x,y}-\psi_x-\psi_y$ Poincar\'e section.  In
each set, we cast 64 initial coordinates uniformly distributed on the
$\psi_x-\psi_y$ plane. For every set of sextupole configuration, we
calculated the new action $r_{x,y}$ for all of the 3 sets of particles,
using the formula $w_0=u_0\mathbf{Z}_0$. For each set, the nonlinearity
measure from Eq.~\eqref{variation} was the optimization objective. In
order to control the distortion for different sets simultaneously, we
adopted the multi-objective genetic algorithm (MOGA)~\cite{Deb:2002}. The
choice of initial values was not unique. The question about how many sets
should be used, and how many points should be cast inside each set is open
for future exploration. After 85 generations and an evolution of 4000
populations, the optimizer converged to an optimal solution, which we
labeled as Solution B in the following section.

Then we compared two solutions, A and B. Solution A is obtained by a
conventional method - minimizing 8 first-order and 23 second-order
nonlinear driving-terms, including
amplitude-dependent-tune-shift~\cite{Li:2016,Chao:2002,Wang:2012}.
Solution B was obtained using the square matrix method as outlined
above. FIG.~\ref{flat_plane} shows that the square matrix method can
significantly reduce the $r=|w_0|$ distortions from a constant at a given
initial amplitude $x=20mm$ and $y=3mm$. As expected, we also observed that
the trajectories of Solution B are much more linear than those of Solution
A in the phase space (FIG.~\ref{phase_y}, top). The spectral analysis
(FIG.~\ref{phase_y}, bottom) indicates that the motion in the case of
Solution B is mainly dominated by a single frequency.

\begin{figure}[!ht]
\centering \includegraphics[width=.7\columnwidth]{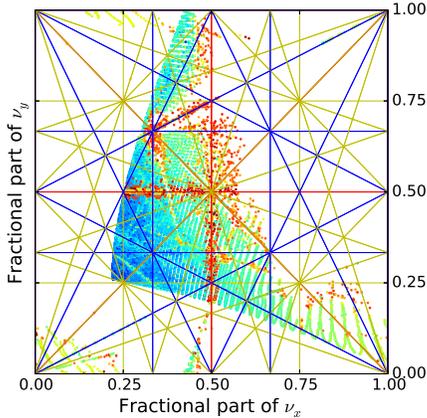}
\caption{\label{fma}Tune footprint for on-momentum dynamic aperture of
  Solution B. The color of each dot represents its tune diffusion
  $log_{10}\sqrt{\Delta\nu_x^2+\Delta\nu_y^2}$ as defined in frequency map
  analysis~\cite{Laskar:2003,Robin:2000}. When the tune crosses the third
  order resonance $3\nu_x=n$, there is no beam loss, and even no obvious
  diffusion.}
\end{figure}

\begin{figure}[!ht]
\centering \includegraphics[width=1.\columnwidth]{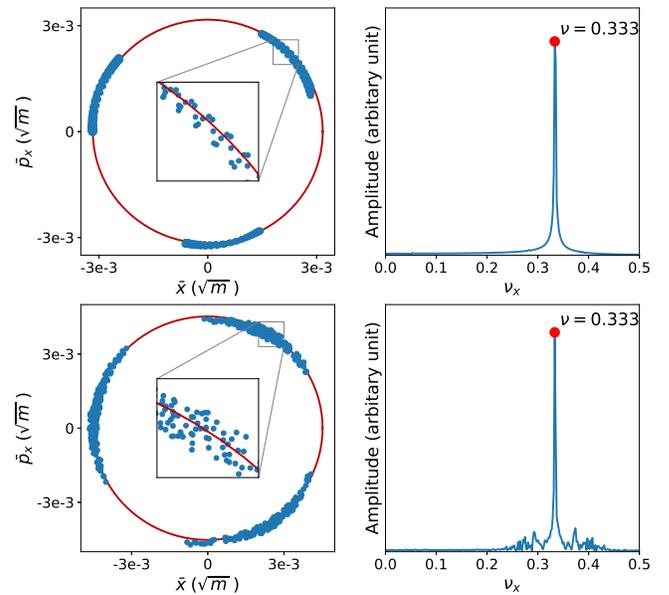}
\caption{\label{third} Simulated horizontal phase space trajectories with
  their tunes at a third order resonance. In the left plots and their
  zoomed-in subplots, the red lines represent constant linear actions.
  The blue dots are the simulated turn-by-turn data. The frequency
  spectrums (right) indicates that the particle can stably stay at the
  resonance line. The top plots are for an ideal machine, and the bottom
  plots for the machine with errors.}
\end{figure}

Here we note that the tune footprint (see FIG.~\ref{fma}) of Solution B
has very large amplitude-dependent tune shift in both planes. It is
remarkable that many particles can survive on a number of resonances at
large amplitudes. FIG.~\ref{third} illustrates a simulated horizontal
trajectory in phase space while its horizontal tune is almost exactly at a
third order resonance. This indicates the irregularity near the resonance
$3\nu_x=n$, has been almost completely eliminated. Usually $3\nu_x$ is
regarded as a dangerous resonance in a sextupole-dominated nonlinear
lattice. For some machines, tunes can cross it at small amplitudes with no
beam loss. When a particle's tune approaches the resonance, its amplitude
will be blown-up and its tune is shifted off the resonance, which serves
as a stability mechanism. The nonlinear force drives particles' tunes and
amplitudes to vary, which leads to a visible tune diffusion and amplitude
fluctuation~\cite{Laskar:2003}. In this case, the stop-band width is wide,
and the motion stability is sensitive to errors. It is difficult for
particles to cross the resonance at large amplitudes. In the past, the
convention was to confine the tune footprint within a narrow range. The
behavior of solution B, however, is very different than Solution
A. FIG. ~\ref{third} illustrates that one particle can stably stay at the
$3\nu_x$ resonance without obvious tune diffusion and amplitude
fluctuation at a large amplitude around $x=13.5mm$. For each trajectory an
unique tune is determined by its amplitude, but not the phase angle. Its
nonlinear behavior is like a near-integrable system. Further exploration
to understand nonlinear dynamic behavior in the vicinity of resonances is
still under way.

\begin{figure}[!ht]
\centering \includegraphics[width=1.\columnwidth]{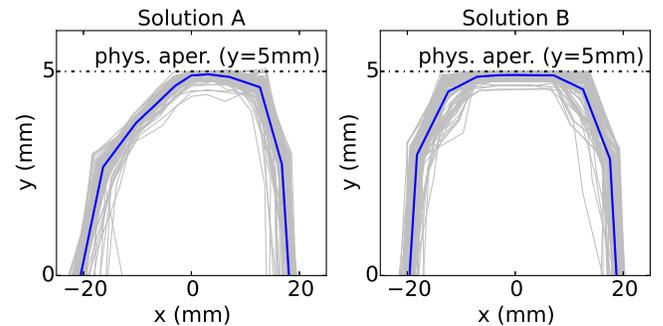}
\caption{\label{physaper} Comparison of effective dynamic apertures of two
  Solutions when a physical aperture limitation $y=5mm$ and multipole
  errors are present. The blue lines are the aperture averaged over 80
  random seeds (light-gray lines). The top-left corner of Solution A's
  aperture is scraped due to the chaotic vertical motion.}
\end{figure}

Our simulation shows that Solution B is quite tolerant to magnet
imperfections. After the specified systematic and random multipole errors
(the typical multipole components normalized to the main components
evaluated at a $25mm$ radius is around the order of $1-3\times 10^{-4}$
~\cite{Skaritka:2010}), and some physical apertures limitation are
introduced into the tracking simulation, the dynamic aperture remains
sufficient for off-axis injection (see FIG.~\ref{physaper}, right). In
particular, particles can still cross the resonance $3\nu_x=n$ smoothly,
and the cancellation of resonance is well preserved (FIG.~\ref{third},
bottom). Experimentally, under this sextupole configuration, 100\%
off-axis injection efficiency into the NSLS-II ring has been achieved,
which is consistent with our analysis and simulation.

It is worth noting that when tight physical apertures are present in the
storage ring, particles with a chaotic motion can be scraped by the
boundary of the physical apertures, which results in a reduction of
effective dynamic aperture. Regular motion is not limited in this way as
can be seen in FIG.~\ref{physaper}.

Over decades, we have followed a common convention -- choosing the
fractional tunes far away from low-order resonances. With this design,
sextupoles are tuned to confine amplitude-dependent tune-shift in order to
avoid crossing them. The solution obtained with our method, however,
obviously violates this convention. This indicates that confining tune
footprint in order to avoid resonance line-crossing is not absolutely
necessary in lattice optimization. Our method suggests a new lattice
design philosophy, where instead of confining tune footprint excursion,
one can tune sextupoles to minimize the variation of $r_{x,y}=|w_{x,y}|$
at different amplitudes to optimize dynamic aperture.

\section{conclusion}
Through the use of linear algebra techniques we developed a new method for
analyzing periodic, nonlinear dynamical systems. Applying Jordan
decomposition to the eigenspace of the square matrix, we found a set of
accurate action-angle variables. The distortions from the flat planes
after mapping constant linear actions to the new actions are one measure
of nonlinearity. Several other measures, such as the two measures given by
Eq.~\eqref{stability} could be used in future exploration. Our method was
successfully field-tested by optimizing the NSLS-II lattice. Most
importantly, optimization using our square matrix method has generated an
unprecedented nonlinear lattice which allows particles to stay exactly on
resonance. Thus the new approach allows relaxed tune footprint, and widely
opened a potential new direction for the search of larger dynamic
aperture. It also provides a different perspective to guide the
understanding of the nonlinear dynamics. The square matrix method is
general and can be applied to other nonlinear dynamical systems with
periodic structure, such as celestial mechanics.

\begin{acknowledgments}
We would like to thank Dr. Y. Hao for sharing his TPSA code, Dr. B. Nash
for the collaboration during the early stage of developing this method,
Dr. M. Borland and Dr. Y.P. Sun for discussion and collaboration on
applying this method to the APS-U ring, Dr. X. Huang for implementing
experimental studies on the SPEAR3 ring, Prof. A. Chao for a stimulating
discussion, and Mr. R. Rainer for editing the manuscript. This work was
supported by Department of Energy Contract No. DE-AC02-98CH10886 and
DE-SC0012704.
\end{acknowledgments}

\bibliography{sqmat_ref}

\end{document}